\documentclass{aa}
\usepackage{txfonts}
\usepackage{graphicx}
\usepackage{natbib}
\bibpunct{(}{)}{;}{a}{}{,}
\def\new{\relax}
\def\newer{\relax}
\begin{document}
\title{Orbits and Masses in the T Tauri System%
\thanks{Based on observations collected at the European Southern
  Observatory, Chile, proposals number 070.C-0162, 072.C-0593,
  074.C-0699, 074.C-0396, 078.C-0386, and 380.C-0179}}
\author{Rainer K\"ohler\inst{1,2}
	\and
	Thorsten Ratzka\inst{3}
	\and
	T.\ M.\ Herbst\inst{2}
	\and
	Markus Kasper\inst{4}
}
\offprints{Rainer K\"ohler, \email{r.koehler@lsw.uni-heidelberg.de}}
\institute{%
	ZAH, Landessternwarte,
	K\"onigstuhl, 69117 Heidelberg, Germany
\and
	Max-Planck-Institut f\"ur Astronomie,
	K\"onigstuhl 17, 69117 Heidelberg, Germany
\and
	Astrophysikalisches Institut Potsdam,
	An der Sternwarte 16, 14482 Potsdam, Germany
\and
	European Southern Observatory,
	Karl-Schwarzschild-Str.\ 2, 85748 Garching bei M\"unchen, Germany
}
\date{Received December 19, 2007 / Accepted February 18, 2008}
%%
%%%%%%%%%%%%%%%%%%%%%%%%%%%%%%%%%%%%%%%%%%%%%%%%%%%%%%%%%%%%%%%%%%%%%%%%%%%%%
%%
\abstract{}
{We investigate the binary star T~Tauri South, presenting the orbital
  parameters of the two components and their individual masses.}
{We combined astrometric positions from the literature with previously
  unpublished VLT observations.  Model fits yield the orbital
  elements of T Tau Sa and Sb.  We use T Tau N as an astrometric
  reference to derive an estimate for the mass ratio of Sa and Sb.}
{Although most of the orbital parameters are not well constrained,
  {\new it is unlikely that T Tau Sb is on a highly elliptical orbit
    or escaping from the system}.  The total mass of T Tau S is rather
  well constrained to $3.0^{+0.15}_{-0.24}\rm\,M_\odot$.  The mass
  ratio Sb:Sa is about 0.4, corresponding to individual masses of
  $M_{\rm Sa} = 2.1\pm0.2\,M_\odot$ and
  $M_{\rm Sb} = 0.8\pm0.1\,M_\odot$.  This confirms that
  the infrared companion in the T Tauri system is a pair of young
  stars obscured by circumstellar material.}
{}
\keywords{Stars: pre-main-sequence --
	  Stars: individual: T Tauri --
          Stars: fundamental parameters --
	  Binaries: close --
          Astrometry --
	  Celestial Mechanics
}
\maketitle
%
%%%%%%%%%%%%%%%%%%%%%%%%%%%%%%%%%%%%%%%%%%%%%%%%%%%%%%%%%%%%%%%%%%%%%%%%%%%%%

\section{Introduction}

T~Tauri was discovered in 1852 by Hind during observations of the
bright variable nebula NGC~1555 (Hind's nebula) \citep{bertout84}.
Almost a century later, \cite{joy45} defined a new class of variables,
which contained eleven stars initially. The members of the class were
later identified as low-mass pre-main-sequence objects.
Joy named these sources `T~Tauri variables', because T~Tau not only
shows the typical features, but also is among the brightest and best
known objects of this group.

However, as observational techniques improved, the character of
T~Tau turned out to be more peculiar than prototypical. One of the
most surprising findings was made by \cite{dyck82}. They observed
T~Tau using one-dimensional speckle interferometry and found an
infrared companion with a separation of less than one arcsecond.
However, neither the relative position of the two sources nor the
identification of the visually bright component could be
derived. Radio observations \citep{cohen82} and subsequent astrometric
comparisons between the radio and the optical data revealed that the
infrared companion was located $0\farcs7\pm0\farcs1$
\citep{devegt82} or $0\farcs6\pm0\farcs1$ \citep{hanson83} south of
the visually bright component. The southern component dominates the
system longward of the near-infared wavelength regime, but no optical
counterpart could be identified down to 19.6\,mag in the V-band
\citep{stapelfeldt98}.

Explanations of the physical nature of the enigmatic infrared
companion ranged from a true protostar still embedded in its
contracting envelope \citep{bertout83}, to a FUOr-like source with the
observed rapid brightening due to accretion \citep{ghez91}, to a
normal T~Tauri star coeval with its primary. In the latter case, the
star has to be obscured either due to a special viewing geometry
through the circumstellar material \citep{calvet94} or enhanced
accretion \citep{koresko97}.

Considerable observational and theoretical effort has been expanded
to understand this source.  Perhaps the most important ingredient in
solving this puzzle came surprisingly: the resolution of
the southern companion into two sources using speckle holography
\citep{koresko00}.
With a separation of ${}\sim0\farcs05$, the two components were named
T~Tau~Sa and T~Tau~Sb. T~Tau~Sb has been identified as a heavily
extincted, actively accreting pre-main-sequence star with a spectral
type M1 \citep{duchene02}, while the spectrum of T~Tau~Sa is
featureless except for a strong Br$\gamma$ emission line, leaving the
nature of the object uncertain.

The northern of the two radio sources found by \cite{schwartz84} is
easy to identify with T~Tau~N, but it was not clear which of the two
components of T~Tau~S should be identified with the southern radio
source. \cite{johnston03} and \cite{loinard03, loinard05} associated
the radio source with T~Tau~Sb.  This would require that the orbital
motion of T~Tau~Sb around T~Tau~Sa {\new shows a dramatic change
  between 1995 and 1998, probably caused by the ejection of Sb from
  the T~Tau~S system \citep{loinard03}.  However, \cite{furlan03}
  found that T~Tau~Sb and the radio source have distinct paths, and
  suggested a fourth object as the source of the southern radio
  emission.}
After further refinement of the picture, it has been concluded that
the radio source may be connected with, but not identical to the
infrared source \citep{johnston04b, johnston04a, loinard07a}.

{\new For completeness, we note that the detection of another source
  $0\farcs27$ north of T~Tau~N has been reported by
  \cite{nisenson85}. They identified a second component in the visual
  wavelength regime, which was later detected in the K-band by
  \cite{maihara91}.  However, this source was not detected by
  \cite{gorham92} nor by \cite{stapelfeldt98}.
  If the additional source exists, probable explanations for the
  non-detections are that the source is variable either intrinsically
  or due to varying extinction. The latter is not unlikely considering
  the complex environment of T~Tau \citep{solf99, herbst07}.}

The presence of the companion Sb at such a small separation offers the
possibility to trace the orbital motion of T~Tau Sa and Sb around each
other in a reasonable time and with no influence of T~Tau N in first
order approximation. This can lead to the determination of the
dynamical mass of T~Tau~S and -- with the help of the common orbit of
the southern pair around the northern component -- the individual
masses of T~Tau~Sa and Sb.  Such information is essential to judge the
different models.  In addition, evolutionary models of
pre-main-sequence objects may benefit from the direct measurement of
stellar masses, since only a few young stellar objects have masses
known independent of theoretical assumptions.

\cite{duchene06} made such a dynamical analysis of the T~Tau~S system
using measurements in both the near-infrared and radio.  A
simultaneous fit of the Keplerian orbit of T~Tau~Sb around T~Tau~Sa
and the orbital motion of T~Tau~Sb around T~Tau~N lead to a dynamical
mass of $3.34\pm0.36\,\rm{M}_{\odot}$ for the southern binary and a
mass ratio Sb:Sa of $0.18\pm0.04$.  This result is consistent with
earlier determinations of the individual masses
\citep{johnston04b, johnston04a}.  The improved accuracy allows us
to conclude that T~Tau~Sa with $2.73\pm 0.31\,\rm{M}_{\odot}$ may
be a Herbig~Ae star whose light is extincted by a highly inclined
circumstellar disc \citep{kasper02,duchene05,duchene06}.
The derived mass of T~Tau~Sb is $0.61\pm 0.17\,\rm{M}_{\odot}$,
in good agreement with its early M-type spectrum.
\cite{duchene06} modeled the orbital motion of T~Tau~S around N by
polynomials in $x$ and $y$.  This does not allow a determination of
the system mass, and prevents the derivation of the dynamical mass of
T~Tau~N.

In this paper, we further refine our knowledge of the orbits and thus
of the individual masses.
We restrict ourselves to measurements made in the near-infrared,
since the identification of the radio sources with T~Tau~Sb is
uncertain \citep{loinard07b}.
{\new A further complication comes from the fact that T~Tau~Sa has not
  been detected in radio observations.  The position of Sb relative to
  Sa can only be computed indirectly, using the position of T~Tau~N and
  the estimated parameters of the orbits.}

We also explore the confidence limits for the resulting orbital
elements, a crucial step when interpreting such highly degenerate
orbits as those in the T~Tau triple system.

The data from the literature, as well as the reduction and analysis of
our own measurements, are presented in Section~\ref{data}. In
Section~\ref{fit}, the fit of the orbital elements and the
determination of the mass ratio is explained. The reliability of the
fit and its implications for our current knowledge of the T~Tau system
is discussed in Section~\ref{discuss}. In Section~\ref{sum} we summarize
our results.

%%%%%%%%%%%%%%%%%%%%%%%%%%%%%%%%%%%%%%%%%%%%%%%%%%%%%%%%%%%%%%%%%%%%%%%%%%%%%

\section{Observations\label{data}}

\subsection{Literature}

T~Tauri has been observed in the near-infrared with various telescopes
and instruments since the discovery of its binary nature. Most of the
measurements used in this work thus were taken from the literature
(see Tables~\ref{ObsNSTab} and \ref{ObsNSabTab}).
The tables include all statistical and
systematic errors reported by the authors in the papers or in
follow-up studies. The data of the observations that did not
resolve the southern binary are listed in Table~\ref{ObsNSTab}.
Table~\ref{ObsNSabTab} lists observations that did resolve T~Tau S and
gives the positions of component Sa relative to N and Sb,
respectively.  \cite{duchene06} give only positions of Sb relative to
N and Sa; the position of Sa relative to N has been computed by us,
taking the propagation of errors into account.

%% Table1.tex
%% created by ./measurements.pl on Tue Aug 28 18:08:00 2007
%% fine-tuned by hand
%% Last Change: Mon Sep  3 17:08:44 2007
\begin{table}
\caption{Astrometric Measurements of T~Tau N -- S (S unresolved)}
\label{ObsNSTab}
\begin{tabular}{ll@{${}\pm{}$}rl@{${}\pm{}$}ll}
\noalign{\vskip1pt\hrule\vskip1pt}
Date (UT) 	& \multicolumn{2}{c}{$d$ [mas]}	& \multicolumn{2}{c}{P.A.~$[^\circ]$} & Source \\
\noalign{\vskip1pt\hrule\vskip1pt}
1989 Dec 10	& $720$&$14$ 	& $175.5$&$1.0$ 	& \cite{ghez95}\\
1990 Nov 10	& $716$&$14$ 	& $176.2$&$1.0$ 	& \cite{ghez95}\\
1991 Nov 19	& $702$&$14$ 	& $176.4$&$1.0$ 	& \cite{ghez95}\\
1992 Feb 20	& $710$&$14$ 	& $176.7$&$1.0$ 	& \cite{ghez95}\\
1992 Oct 11	& $692$&$15$ 	& $177.0$&$1.4$ 	& \cite{ghez95}\\
1993 Nov 25	& $701$&$14$ 	& $177.0$&$1.0$ 	& \cite{ghez95}\\
1993 Dec 26	& $690$&$14$ 	& $176.0$&$1.1$ 	& \cite{ghez95}\\
1994 Sep 22	& $689$&$14$ 	& $175.9$&$1.2$ 	& \cite{ghez95}\\
1994 Oct 19	& $689$&$14$ 	& $178.2$&$1.1$ 	& \cite{ghez95}\\
1994 Dec 25	& $718$&$ 9$ 	& $177.6$&$0.7$ 	& \cite{roddier00}\\
1997 Nov 17	& $709$&$ 6$ 	& $180.2$&$0.5$ 	& \cite{roddier00}\\
1998 Nov ~2	& $704$&$ 6$ 	& $180.7$&$0.5$ 	& \cite{roddier00}\\
1999 Nov 23	& $699$&$ 6$ 	& $181.0$&$0.5$ 	& \cite{roddier00}\\
1997 Dec ~4	& $685$&$13$ 	& $179.5$&$1.0$ 	& \cite{white01}\\
1997 Dec ~6	& $698$&$13$ 	& $179.1$&$1.0$ 	& \cite{white01}\\
2000 Feb 20	& $691$&$ 2$ 	& $181.7$&$0.8$ 	& This work\\
\noalign{\vskip1pt\hrule\vskip1pt}
\end{tabular}
\end{table}

%% Table4.tex
%% created by ./measurements.pl on Tue Aug 28 18:08:00 2007
%% fine-tuned by hand
%% Last Change: Mon Nov  5 14:59:29 2007
\begin{table*}
\def\1{\phantom{1}}

\def\fnote#1{^{\mathrm{#1}}}
\caption{Astrometric Measurements of T~Tau N -- Sa -- Sb}
\label{ObsNSabTab}
%            date|sep       esepPA         ePa|sep       esepPA         ePA|src
\begin{tabular}{lcl@{${}\pm{}$}ll@{${}\pm{}$}lcl@{${}\pm{}$}ll@{${}\pm{}$}lcl}
\noalign{\vskip1pt\hrule\vskip1pt}
Date (UT) 	&& \multicolumn{4}{c}{T Tau N - Sa}		&& \multicolumn{4}{c}{T Tau Sa - Sb} 	&& Source \\
          	&& \multicolumn{2}{c}{$d$ [mas]} & \multicolumn{2}{c}{P.A.~$[^\circ]$}
								&& \multicolumn{2}{c}{$d$ [mas]} & \multicolumn{2}{c}{P.A.~$[^\circ]$} \\
\noalign{\vskip1pt\hrule\vskip1pt}
1997 Oct 12 	&& $693  $&$12$	 	& $178.6$&$0.9$ 	&& $\1 51$  &$9  $ & $218  $&$8  $ 	&& \cite{duchene06}\\
1997 Dec 15 	&&\multicolumn{2}{c}{--}& \multicolumn{2}{c}{--}&& $\1 53$  &$9  $ & $225  $&$8  $ 	&& \cite{koresko00}\\
2000 Feb 20 	&& $686.3$&$2.4$	& $180.5$&$0.8$ 	&& $\1 79$  &$2  $ & $253  $&$2  $ 	&& This work / \cite{koehler00a}\\
2000 Nov 19 	&& $702  $&$5  $	& $180.8$&$0.2\fnote{a}$&& $\1 92$  &$3  $ & $268.1$&$1.6\fnote{a}$&& \cite{duchene02}\\
%2002 Oct 30 	&& $697  $&$2  $	& $183.1$&$0.3$ 	&& $107  $&$4  $ & $283.4$&$2.1$ 	&& \cite{beck04} --- SAME AS Schaefer\\
2002 Oct 30 	&& $697.6$&$2.1$	& $183.3$&$0.3$ 	&& $106.5$&$2.6$ & $284.5$&$1.4$ 	&& \cite{schaefer06}$\fnote{b}$\\
2002 Nov 20 	&& $693  $&$4  $	& $183  $&$1  $ 	&& $\1 94.3$&$3.3$ & $278  $&$1  $ 	&& \cite{mayama06}\\
2002 Dec 13 	&& $695  $&$7  $	& $183.3$&$0.7$ 	&& $108  $&$1  $ & $284.9$&$0.9$ 	&& \cite{duchene05}\\
2002 Dec 15 	&& $693.5$&$1.4$	& $182.3$&$0.3$ 	&& $106.4$&$1.6$ & $283.9$&$0.4$ 	&& This work\\
2002 Dec 20 	&&\multicolumn{2}{c}{--}& \multicolumn{2}{c}{--}&& $110  $&$4  $ & $282  $&$2.1$ 	&& \cite{beck04}\\
2002 Dec 24 	&& $691  $&$12 $	& $181.7$&$1.1$ 	&& $108  $&$5  $ & $287.8$&$2.6$ 	&& \cite{furlan03}\\
2003 Nov 18 	&& $700.4$&$2.9$	& $183.2$&$0.6$ 	&& $116.5$&$3.1$ & $291  $&$1.6$ 	&& \cite{schaefer06}\\
2003 Dec 12 	&& $698  $&$5  $	& $181.9$&$1.2$ 	&& $118  $&$2  $ & $288.6$&$1.1$ 	&& \cite{duchene05}\\
2003 Dec 12 	&& $696.2$&$2.0$	& $182.8$&$0.1$ 	&& $113.3$&$1.0$ & $288.5$&$0.8$ 	&& This work\\
%2004 Nov ~3 	&&\multicolumn{2}{c}{--}& \multicolumn{2}{c}{--}&& $124.3$&$7.6$ & $299.6$&$5.2$ 	&& \cite{ratzka07}\\ %% MIDI
2004 Nov 23 	&& $690  $&$3  $	& $186  $&$1  $ 	&& $100  $&$2  $ & $298  $&$1  $ 	&& \cite{mayama06}\\
2004 Dec ~9 	&& $693.2$&$1.2$	& $183.5$&$0.1$ 	&& $118.9$&$1.1$ & $295.9$&$0.6$ 	&& This work\\
2004 Dec 19 	&& $689  $&$3  $	& $183.6$&$0.4$ 	&& $116  $&$4  $ & $294.1$&$1.4$ 	&& \cite{duchene06}\\
2004 Dec 24 	&& $698.5$&$2.2$	& $184.7$&$0.3$ 	&& $119.6$&$2.8$ & $297.2$&$1.4$ 	&& \cite{schaefer06}\\
2005 Feb ~9 	&& $704.0$&$1.6$	& $184.4$&$0.1$ 	&& $120.7$&$1.9$ & $300.4$&$0.9$ 	&& \cite{schaefer06}\\
2005 Mar ~9 	&& $702.8$&$2.4$	& $184.4$&$0.2$ 	&& $120.3$&$2.9$ & $300.6$&$1.4$ 	&& \cite{schaefer06}\\
2005 Mar 24 	&& $703.7$&$0.9$	& $184.3$&$0.1$ 	&& $121.5$&$1.2$ & $300.4$&$0.6$ 	&& \cite{schaefer06}\\
2005 Oct 20 	&& $703.3$&$1.2$	& $184.9$&$0.1\fnote{c}$&& $123.7$&$1.3$ & $303.3$&$0.6$ 	&& \cite{schaefer06}\\
2005 Nov 13 	&& $694  $&$2$		& $184.4$&$0.2$ 	&& $119  $&$1  $ & $300.6$&$1.0$ 	&& \cite{duchene06}\\
2005 Dec ~6 	&& $703.0$&$5.7$	& $185.8$&$0.5$ 	&& $121.9$&$6.8$ & $304.8$&$3.2$ 	&& \cite{schaefer06}\\
2006 Oct 11 	&& $694.4$&$0.9$	& $184.9$&$1.0$ 	&& $125.1$&$0.7$ & $304.7$&$1.0$ 	&& This work\\
2007 Sep 16	&& $693.6$&$1.8$	& $185.0$&$0.2$		&& $127.0$&$0.5$ & $308.2$&$0.3$	&& This work\\
\noalign{\vskip1pt\hrule\vskip1pt}
\end{tabular}
\begin{list}{}{}
  \item[$^{\mathrm{a}}$] Corrected value according to \cite{duchene06}
  \item[$^{\mathrm{b}}$] This is a re-analysis of the data presented
    in \cite{beck04} using a more sophisticated PSF fitting routine.
  \item[$^{\mathrm{c}}$] $90^\circ$ subtracted from the position angle
    in \cite{schaefer06} to correct for the orientation of the camera
    (Schaefer, priv.\ comm.).
\end{list}
\end{table*}

\subsection{ALFA}

\cite{koehler00a} presented results of observations obtained in
February 2000 with the adaptive optics system ALFA at the
3.5\,m telescope on Calar Alto, Spain.  The $H$ photometric band was
used to take advantage of the smaller diffraction limit compared to
the $K$ band.  T~Tau N served as PSF-reference to deconvolve T~Tau~S
using speckle interferometric techniques.  This allowed them to
resolve the Sa-Sb binary, and confirm the discovery of
\cite{koresko00}.

\begin{figure}[ht]
\centerline{\includegraphics[width=\hsize]{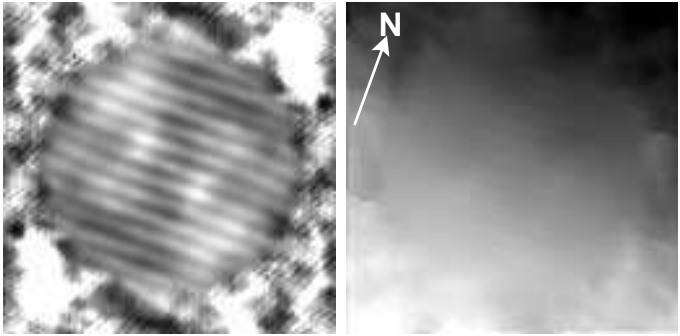}}
\caption{Result of the speckle-observation of T~Tauri on February 20,
  2000.  Shown are modulus (left) and phase (right) of the complex
  visibility (the Fourier-transform of the image).  The N-S binary
  pair causes the fringe pattern in the modulus and the steps in the
  phase.  The fringe pattern due to the binary nature of T~Tau~S is
  too weak to be recognizable.}
\label{NSvisi}
\end{figure}

We revisited these observational data, in order to measure the
position of component Sa relative to N.  In our new analysis, we used
observations of the star PPM\,119798, taken shortly after the images
of T~Tauri, to deconvolve the triple system. We again applied
our own speckle program; for a detailed description see, e.g.,
\cite{koehler00b}.
The resulting visibility appears in Fig.~\ref{NSvisi}. It clearly
shows the characteristic fringe pattern of the N-S binary.
However, the signal of the Sa-Sb pair is obscured by random
fluctuations.  One has to keep in mind that T~Tau N contributes more
than 90\,\% of the flux of the system in the $H$ band, resulting in
correspondingly high speckle noise. Therefore, it was not possible to
measure the relative position of T~Tau N and Sa, but only the position
of the combined light of Sa and Sb relative to N.  The latter is
listed in Table~\ref{ObsNSTab}. We then computed the offset of T Tau
Sa from the center of light of Sa and Sb, using the known position of
Sa relative to Sb and the flux ratio of Sb/Sa of $0.259\pm0.011$
\citep{koehler00a}.  The results are given in Table~\ref{ObsNSabTab}.

\subsection{NAOS/CONICA}

We also report on observations with NAOS/CONICA
(NACO for short), the adaptive optics, near-infrared camera at the ESO
Very Large Telescope on Cerro Paranal, Chile \citep{Rousset03, lenzen03}.
They are marked ``this work'' in Table~\ref{ObsNSabTab}.
The NACO observations in 2002 and 2003 were carried out in the course
of the NACO Guaranteed Time Observations (PI Tom
Herbst\footnote{\cite{herbst07} give astrometric data derived from the
  observations in 2002.  However, their analysis was carried out
  independently from ours.}), while the observations in 2004, 2006,
and 2007 were regular proposals for open time.  We use only imaging
observations in the $K$ or $K_s$ photometric band for the orbit
determination.

\begin{figure}[ht]
\centerline{\includegraphics[width=\hsize]{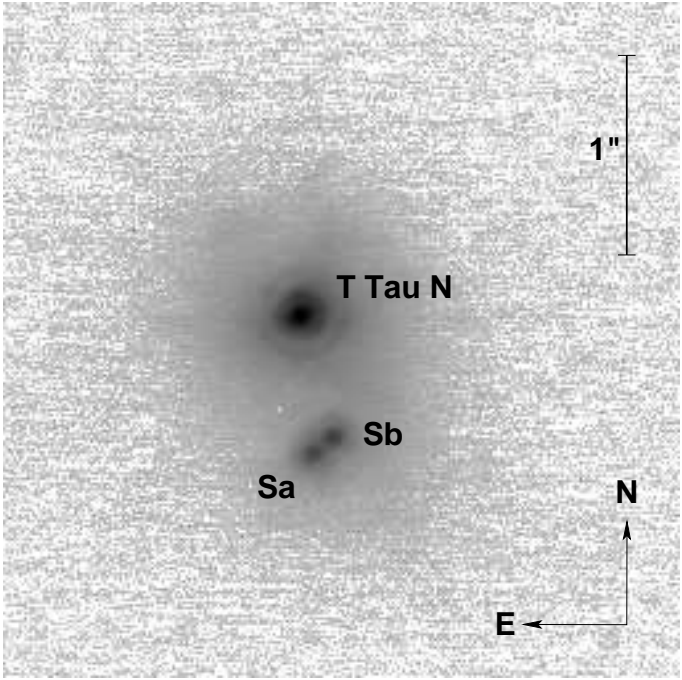}}
\caption{Image of the T Tau system taken with NACO in the $K_s$
  photometric band on 2006 Oct 11, shown with a logarithmic scale.}
\label{NACOpic}
\end{figure}

The NACO images were sky subtracted with a median sky image, and
bad pixels were replaced by the median of the closest good neighbors.
Finally, the images were visually inspected for any artifacts or
residuals.  Figure~\ref{NACOpic} shows an example of the results.

The Starfinder program \citep{Diolaiti00} was used to measure the
positions of the stars.  The positions in several images taken during
one observation were averaged, and their standard deviation used to
estimate the errors.  To derive the exact pixel scale and
orientation of the detector, we took images of fields in the Orion
Trapezium during each observing campaign.  The measured positions
of the stars were compared with the coordinates given in
\cite{mccaughrean94} by the astrometric software ASTROM\footnote{see
http://www.starlink.rl.ac.uk/star/docs/sun5.htx/sun5.html}.  The
calibrated separations and position angles appear in
Table~\ref{ObsNSabTab}.

%%
%% 2002: 070.C-0162, Conica GTO, Herbst
%% 2003: 072.C-0593, Conica GTO, Herbst
%% 2004: 074.C-0699, Kasper/Herbst/Calvet: Fabry-Perot, not used
%% 2004: 074.C-0396, Ratzka/Leinert/K\"ohler
%% 2005: no NACO obs
%% 2006: 078.C-0386, Ratzka/Koehler/Leinert
%% 2007: 380.C-0179, RKL
%%

%%%%%%%%%%%%%%%%%%%%%%%%%%%%%%%%%%%%%%%%%%%%%%%%%%%%%%%%%%%%%%%%%%%%%%%%%%%%%

\section{Orbit Determination\label{fit}}

\subsection{The Orbit of T Tauri Sa/Sb}
\label{SabFitSect}

\begin{figure}[ht]
\centerline{\includegraphics[width=\hsize]{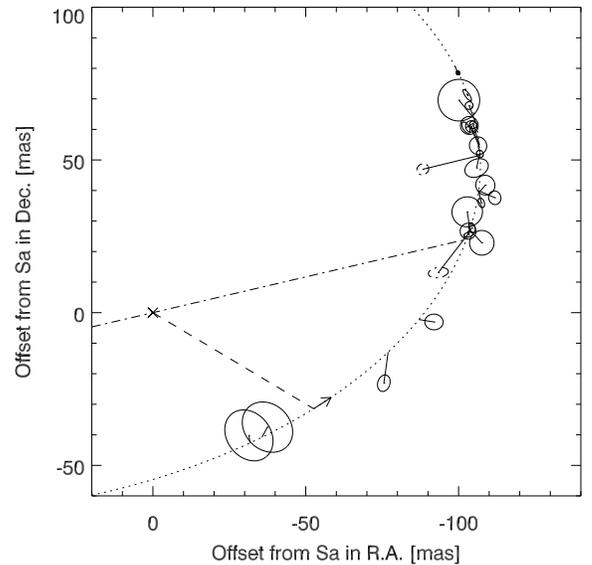}}
\caption{Best fitting orbit of T~Tau~Sb around Sa, in the rest frame
  of Sa. The observed positions are marked by their error ellipses and
  lines connecting the observed and calculated position at the
  time of the observations.  The dash-dotted line indicates the line
  of nodes, the dashed line the periastron, and the arrow shows the
  direction of the orbital motion.  The two dash-dotted ellipses are
  the measurements by \cite{mayama06}.
}
\label{OrbSabPic}
\end{figure}

\begin{figure}[ht]
\centerline{\includegraphics[width=\hsize]{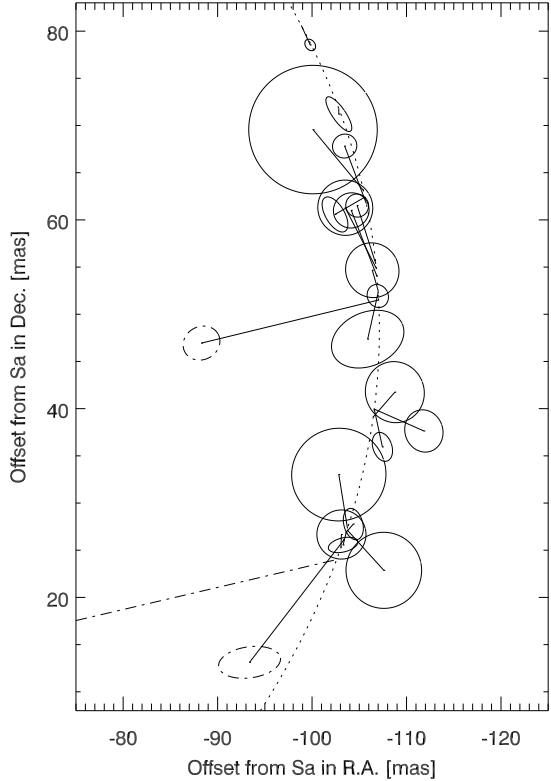}}
\caption{Detailed view of the orbit of T~Tau~Sb in the rest frame of
  Sa, and the observations since 2002.  The symbols are the same as in
  Fig.~\ref{OrbSabPic}.}
\label{OrbSabPicDetail}
\end{figure}

We estimated the orbital parameters of the Sa-Sb pair by fitting orbit
models to all observations listed in Table~\ref{ObsNSabTab}, except the
positions given by \cite{mayama06}.  These two data points show
significant offsets from all the other measurements
(Fig.~\ref{OrbSabPic} and \ref{OrbSabPicDetail}).  Since we do not
know the reason for this discrepancy and have no way to correct it,
we excluded these points from the fit.  To avoid any
systematic errors due to uncertainties of the distance, the separations
are expressed in units of milli-arcseconds within the fitting program.
Only the final result was converted to AU.

To find the orbital fit with the minimum $\chi^2$, we employed a
method similar to that outlined in \cite{Hilditch01}: a grid-search in
eccentricity $e$, period $P$, and time of periastron $T_0$.  At each
grid point, the Thiele-Innes elements were determined by a linear fit
to the observational data using Singular Value Decomposition.  From
the Thiele-Innes elements, the semimajor axis $a$, the angle between
node and periastron $\omega$, the position angle of the line of nodes
$\Omega$, and the inclination $i$ were computed.

We decided to scan a wide range of parameter values:
200 points within $0 \le e \le 2$, 250 points within $10{\rm\,yr} \le
P < 3100{\rm\,yr}$, and initially 100 points for $T_0$ distributed
over one orbital period.  After the initial scan over $T_0$, the best
estimate for $T_0$ was improved by re-scanning a {\newer narrower}
range in $T_0$ centered on the minimum found in the coarser scan.
The range was reduced by a factor of 10, but the number of grid points
was held constant.  This grid refinement was repeated until the step
size was less than one day.

For every point in the $e$-$P$-plane, the set of orbital elements
resulting in the minimum $\chi^2$ was stored, to allow later analysis
of $\chi^2$ as a function of $e$ and $P$.  The
parameter giving the time of periastron $T_0$ is periodic: we obtain
the same orbit if we add one orbital period to $T_0$.  Furthermore, a
change in $T_0$ only shifts the position at a given time within the
orbit.  A value of $T_0$ that has a small offset from the optimal
$T_0$ will therefore shift the expected positions away from the
measured positions and result in a larger $\chi^2$.  For
these reasons, we are confident that our grid search found the
globally optimum $T_0$ and did not store all the sets of orbital
elements for non-optimum $T_0$.

\begin{figure}[ht]
\centerline{\includegraphics[angle=90,width=\hsize]{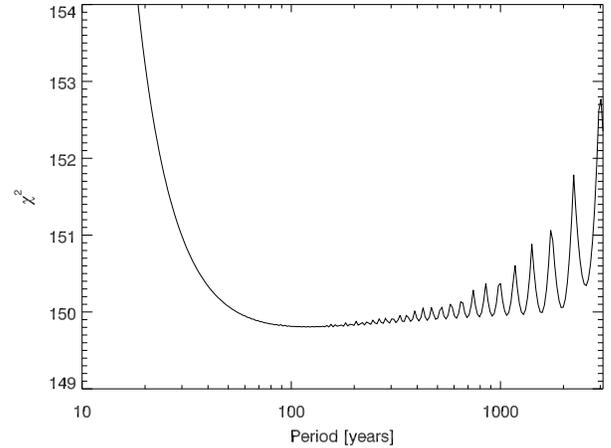}}
\caption{The result of the first step in our orbit-fitting procedure.
  Shown is the minimum $\chi^2$ for a given period, i.e.\ we hold the
  period fixed and search for the optimum set of the remaining 6
  orbital elements.
  Solutions with unphysically high system masses (${}>250\rm\,M_{\sun}$)
  have been excluded.  The ripples at periods
  ${}\protect\ga200$\,years are artefacts caused by the grid search.
  {\new Note that the $\chi^2$ shown here has not been divided
  by the number of degrees of freedom.}
}
\label{chiminfig}
\end{figure}

\begin{figure}[ht]
\centerline{\includegraphics[angle=90,width=\hsize]{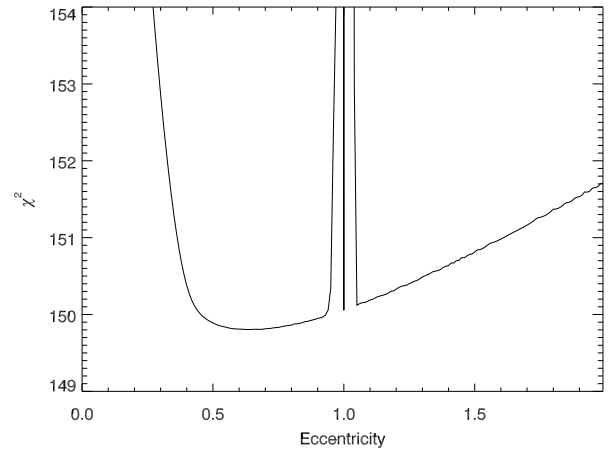}}
\caption{Like Fig.~\ref{chiminfig}, but showing the minimum $\chi^2$
  as function of eccentricity instead of period.  Note that some
  orbits with $e>1$ (i.e.\ unbound orbits) have a $\chi^2$ not much
  higher than the minimum.}
\label{chiminefig}
\end{figure}

\begin{figure}[ht]
\centerline{\includegraphics[angle=90,width=\hsize]{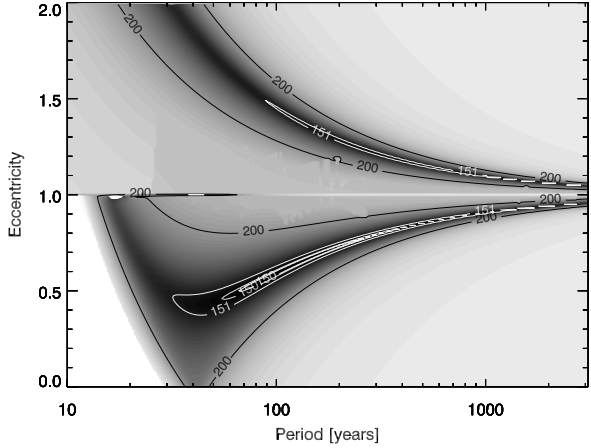}}
\caption{Contour plot of the minimum $\chi^2$ as function of period
  and eccentricity.  The contour $\chi^2=151$ marks the 68\,\%
  confidence intervals for the period or eccentricity (cf.\
  Figs.~\ref{chiminfig} and \ref{chiminefig}).}
\label{chigridfig}
\end{figure}

The resulting $\chi^2$-minimum is very broad
(Figs.~\ref{chiminfig} -- \ref{chigridfig}).  For example,
Fig.~\ref{chiminfig} shows that orbits with periods between 30 and
more than 1000 years are within $\Delta\chi^2 < 1$ of the best orbit
(the $1\sigma$ confidence limit, \citealp{press92}).  The situation for
the eccentricity $e$ is similar (Fig.~\ref{chiminefig}):
Eccentricities between about 0.4 and 0.8, and even some unbound orbits
with $e>1$ are within $\Delta\chi^2 < 1$.

We improved the results of the grid-search with a Levenberg-Marquardt
$\chi^2$ minimization algorithm \citep{press92} that fits for all 7
parameters simultaneously.  The simple-minded approach would be to use
the orbital elements with the minimum $\chi^2$ found with the
grid-search.  However, given the structure of the $\chi^2$-plane
(Fig.~\ref{chigridfig}), there is always the danger that the
algorithm converges on a local instead of the global minimum.  To
avoid this, we decided to use {\em all\/} orbits resulting from the
grid-search as starting points and carried out $250\times200$ runs of
the Levenberg-Marquardt algorithm (the grid spans 250 periods and 200
eccentricities).  The orbit with the globally minimum $\chi^2$ found
in this way is shown in Figs.~\ref{OrbSabPic} and
\ref{OrbSabPicDetail}, and its elements are listed in
Table~\ref{OrbSabTab}.

\subsubsection{Error estimates for the Orbital Elements}

The Levenberg-Marquardt algorithm also computes the covariance matrix
of the fitted orbital parameters, which allows derivation of standard
error estimates for the parameters.  However, the algorithm assumes
that the $\chi^2$ function can be approximated by a quadratic form in
the region close to the minimum.  While this is a reasonable assumption
for iteratively searching the minimum, it is clearly a rather bad
assumption if one is interested in confidence limits for the fitted
parameters (cf.\ Fig.~\ref{chiminfig} and \ref{chiminefig}).  Better
error estimates can be obtained by studying the $\chi^2$ function
around its minimum \citep{press92}.  Since we are interested in the
confidence interval for each parameter taken separately, we have to
perturb one parameter (for example $T_0$) away from the minimum, and
optimize all the other parameters.  Any perturbation of a parameter
will of course lead to a larger $\chi^2$.  The range in $T_0$
within which $\chi^2(T_0) - \chi^2_{\rm min} < 1$ defines the 68\,\%
confidence interval for $T_0$.  This interval is usually not symmetric
around the $T_0$ of the best fit, therefore we list in
Table~\ref{OrbSabTab} separate limits for positive and negative
perturbations.  It should be noted that these limits describe the
parameter range that contain 68\,\% of the probability distribution,
which is equivalent to the commonly used $1\sigma$-errors.  However,
the errors are not normally distributed, therefore a $2\sigma$
interval will {\em not\/} contain 95\,\% of the probability
distribution.

%
% orbit.tex
% created by lmfinderr on Wed Feb 20 17:17:50 2008
%
\begin{table}
\caption{Parameters of the best orbital solution for Sa -- Sb.}
\label{OrbSabTab}
%% stretch a bit so that sub- and superscripts don't overlap
\renewcommand{\arraystretch}{1.3}
\begin{center}
\begin{tabular}{lcr@{}l}
\noalign{\vskip1pt\hrule\vskip1pt}
Orbital Element				&  & \multicolumn{2}{c}{Value} \\
\noalign{\vskip1pt\hrule\vskip1pt}
Date of periastron $T_0$			&  & $2451091$ & $\,^{+ 250}_{-1200}$\\
						&  & (1998 Oct  4)\span\\
Period $P$ (years)				&  & $     93$ & $\,^{+ 115}_{  -55}$\\
Semi-major axis $a$ (mas)			&  & $    201$ & $\,^{+ 430}_{ -110}$\\
Semi-major axis $a$ (AU)			&  & $     30$ & $\,^{+  63}_{  -16}$\\
Eccentricity $e$				&  & $   0.57$ & $\,^{+0.20}_{-0.22}$\\
Argument of periastron $\omega$ ($^\circ$)	&  & $  300.6$ & $\,^{+12.0}_{-30.0}$\\
P.A. of ascending node $\Omega$ ($^\circ$)	&  & $  283.2$ & $\,^{+ 4.0}_{ -7.0}$\strut\\
Inclination $i$ ($^\circ$)			&  & $   55.1$ & $\,^{+ 2.0}_{-20.0}$\\
System mass $M_S$ ($M_\odot$)			&  & $   2.96$ & $\,^{+0.15}_{-0.24}$\\
%%$\chi^2$	       				&  & $  147.9$	\\
reduced $\chi^2$				&  & $    3.8$\\
\noalign{\vskip1pt\hrule\vskip1pt}
Mass ratio $M_{\rm Sb}/M_{\rm Sa}$		&  & $   0.39$ & $\,\pm 0.06$\\
Mass of Sa $M_{\rm Sa}$ ($M_\odot$)		&  & $   2.13$ & $\,^{+0.14}_{-0.20}$\\
Mass of Sb $M_{\rm Sb}$ ($M_\odot$)		&  & $   0.83$ & $\,^{+0.10}_{-0.11}$\\
\noalign{\vskip1pt\hrule\vskip1pt}
\end{tabular}
\end{center}
\end{table}

{\new Orbits with $\chi^2-\chi^2_{\rm min} < 1$ are not necessarily
  all within a single region in parameter space around the global
  minimum.  There may be local minima that also fulfill the
  criterion.  In fact, some of our Levenberg-Marquardt fits converged
  on orbital solutions with eccentricity $e>1$ and
  $\chi^2-\chi^2_{\rm min} < 1$.  These orbits are within the 68\,\%
  confidence limit.  Therefore, we cannot fully exclude that the
  system is unbound, although it is more likely to be bound.}

The system mass $M_S$ given in Table~\ref{OrbSabTab} is computed from
the semi-major axis and the period.  To convert the angular separation
in milli-arcseconds (mas) to a linear separation in AU, a distance to
T~Tau of $146.7\pm0.6\rm\,pc$ was adopted
\citep{loinard07b}\footnote{\newer \cite{loinard07b} give a distance
  of $147.6\pm0.6\rm\,pc$ in the abstract of their paper, but
  $146.7\pm0.6\rm\,pc$ in the main text.  Only the latter is
  compatible with the parallax they report.}.
The semi-major axis $a$ and the period $P$ of the T~Tau~S binary are
correlated around the minimum of the $\chi^2$-surface: If one sets the
period to a value {\newer higher} than at the minimum $\chi^2$ and optimizes
the other 6 parameters, then the resulting $a$ will also be larger
than at the global minimum.  This has to be taken into account in the
computation of the error of the system mass.  One way to do this is to
take into account the covariance of $a$ and $P$.  However, since the
covariance matrix is a good approximation only in a small region
around the $\chi^2$ minimum, we choose a different approach: During
the determination of the confidence intervals for $a$ and $P$, we also
compute the system mass.  The maximum and minimum masses within the
region where $\Delta\chi^2<1$ are adopted as confidence limits and
listed in Table~\ref{OrbSabTab}.  It is worth noting that we did not
have to decide whether we want to use the results from the confidence
interval in $a$ or in $P$ -- the orbits are identical.  This is caused
by the strong correlation between $a$ and $P$, which in turn leads to a
small error in $M_S$ compared to the larger uncertainties in $a$ and
$P$  (see also Sect.~\ref{exactmasssec}).

\subsection{The Orbit of T Tauri S and the Mass Ratio Sb/Sa}
\label{NSFitSect}

\begin{figure}[ht]
\centerline{\includegraphics[width=\hsize]{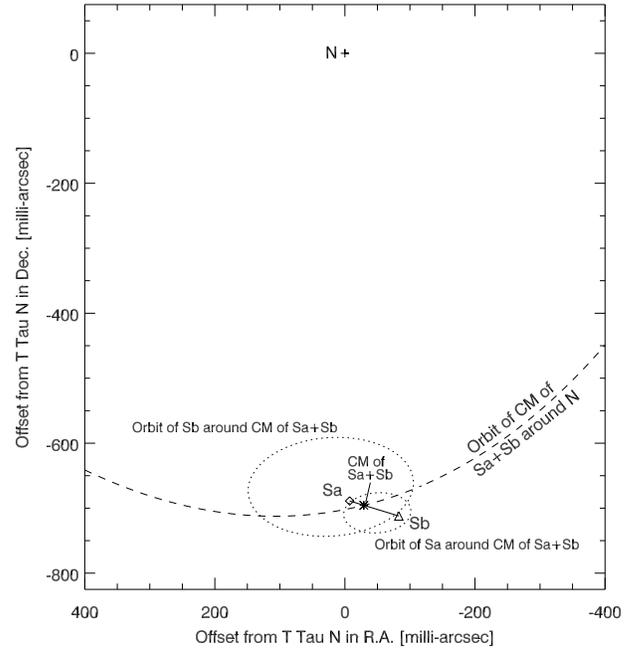}}
\caption{Schematic representation of the stars and orbits in the
  T Tauri system.  The origin of the coordinate system is fixed on T
  Tau N (marked by a plus sign), therefore the center of mass of Sa
  and Sb describes an orbit around N.  The orbit resulting from our
  model fits is indicated by the dashed line, while the position of
  the CM of Sa and Sb on February 20, 2000 is denoted by an asterisk.
  The orbits of Sa and Sb around their CM according to our model fits
  are drawn as dotted lines, the positions of the stars on February
  20, 2000 (computed from the model) are marked by a diamond and a
  triangle.}
\label{geometryfig}
\end{figure}

The positions of T~Tau Sb relative to Sa allow us to determine only the
combined mass of Sa and Sb.  To compute the individual masses, we need
to know the mass ratio $q$, which can be computed if the position of
the center of mass (CM) of Sa and Sb is known.  Unfortunately,
we cannot observe the CM directly.  However, we
know that the CM of Sa and Sb is in orbit around T~Tau~N\footnote{%
  In reality, both N and the center of mass of Sa/b are in orbit
  around the center of mass of the entire system, but this does not
  change the formalism if only relative positions are used.},
and that Sa and Sb are in orbit around their CM (see
Fig.~\ref{geometryfig}).

One might think that in order to derive the mass ratio $q$, it is
necessary to create a model of the entire triple system, and fit both
the orbit of the CM, the orbit of Sa and Sb, and the mass ratio
simultaneously.  However, we know that the CM is always on the line
between Sa and Sb, and its distance from Sa is the constant fraction
$q/(1+q)$ of the separation of Sa and Sb.  The positions of Sa and Sb
are usually observed at the same time as N, so we can use the {\em
  observed\/} separation vectors Sa-Sb instead of a model orbit.

Our model describes the position of the CM of Sa and Sb in two ways:
First, it is on a Keplerian orbit around N, which is described by 7
orbital elements.
Second, the position of the CM can be computed from the positions of
Sa and Sb, and the mass ratio (which is treated as a free parameter).
Standard error propagation is used to obtain an error estimate for
this position.  To compute $\chi^2$, we compare the position of the CM
from the orbit around N with the positions derived from the observations.
Our model has therefore only 8 free parameters, the 7 elements which
describe the orbit of the CM of Sa/b around N, and the parameter
$f=q/(1+q)$.  The parameter $f$ is often called fractional mass
\citep{heintz78}, since it is the secondary star's fraction of the
total mass in a binary.  It is useful in our case because it also
describes the fractional offset of the CM from Sa, i.e.\ the
separation between Sa and the CM divided by the separation between Sa
and Sb.  For a grid-search, $f$ is better suited than $q$,
because $f$ is confined to the range 0 to 1, while $q$ is a number
between 0 and infinity.

The fitting procedure is similar to that used for the orbit of Sa/b,
except that the grid-search is carried out in 4 dimensions:
eccentricity $e$, period $P$, time of periastron $T_0$, and the
fractional mass~$f$.
Singular Value Decomposition was used to fit the
Thiele-Innes constants, which give the remaining orbital elements.
We stress that the orbital elements in this fit describe the orbit of
the N/S binary, only the fractional mass $f$ refers to the pair Sa/b.

We do not expect that the elements of the orbit of S around N will be
well constrained by the small section of the orbit observed so far.
The orbit model is used mainly as a way to smooth out the measured
positions.  \cite{duchene06} used second-order polynomials to do this,
but we believe that it is better to use a physical model, i.e.\ a
Kepler-orbit.  With a physical model, we can check whether the
parameters of the orbit are reasonable, e.g.\ the period, semi-major
axis, and the system mass (see below).  Furthermore, a Kepler-orbit is
fully described by 7 parameters, only one parameter more than the two
polynomials used by \cite{duchene06}.

For the four dimensional grid search, we used 80 points for $f$ in the
range 0 to 0.8, 100 points for $e$ between 0 and 1, 125 points for $P$
in the range 100 to 3100 years, and initially 100 points for $T_0$
distributed uniformly over one orbital period.  Similar to the fit for
the orbit of Sa/Sb, the grid in $T_0$ was refined until the grid
spacing was less than 3 days.  The optimum parameter sets for all $f$,
$e$, and $P$ points were stored, but only the parameter set for the
best $T_0$ point.
The most important insight we gain from this grid search is that the
orbital parameters are not very well constrained, while $\chi^2$ as
a function of $f$ shows a distinct minimum.

Finally, the grid-point with the smallest $\chi^2$ was used as
starting point for a Levenberg-Marquardt fit.  Since this fit method
assumes that the measurement errors are independent of the fitting
parameters, $f$ cannot be treated like the other parameters.  Instead,
we perform a grid search over a {\newer narrow} range in $f$ to find
the minimum.

Fits to small sections of an orbit often result in unphysically high
system masses (several 100\,$M_\odot$).  Such an orbit is clearly not
a good solution, even if its $\chi^2$ is very small.  To prevent the
Levenberg-Marquardt fit from exploring unphysical regions of parameter
space, we added a mass constraint to the computation of $\chi^2$:
$$
\chi^2 = \sum_i \left(\vec r_{i,\rm obs} - \vec r_{i,\rm model}
			\over \Delta\vec r_{i,\rm obs}\right)^2
	+ \left(M_{\rm est}-M_{\rm model} \over\Delta M_{\rm est}\right)^2
$$
Here $\vec r_{i,\rm obs}$ and $\Delta\vec r_{i,\rm obs}$ are the
observed position at time $i$ and the corresponding error, $\vec
r_{i,\rm model}$ is the position at time $i$ computed from the model,
$M_{\rm est}$ and $\Delta M_{\rm est}$ are the estimated system mass
and its error, and $M_{\rm model}$ is the system mass computed from
the orbit model.
We use $M_{\rm est} = 5\rm\,M_\odot$ and $\Delta M_{\rm est} =
1\rm\,M_\odot$.  This is a reasonable estimate for the total mass of
T~Tau~N, Sa, and Sb, based on our results for $M_S$ in
Sect.~\ref{SabFitSect} and the mass of T~Tau~N derived by
\cite{loinard07b}, who estimated it from the spectral energy
distribution.  From comparison with theoretical pre-main-sequence
tracks, they derived a mass of $1.83^{+0.20}_{-0.16}$ or
$2.14^{+0.11}_{-0.10}\rm\,M_\odot$ (depending on the theoretical tracks
used).

An important question is whether we want to include observations that
did not resolve T Tauri S in the fit.  These measurements yield only
the position of the center of {\em light\/} of Sa/b, not the center of
{\em mass}.  The offset between the two centers is unknown and not
even constant, since both components are photometrically variable
\citep{beck04}.  On the other hand, the data obtained before
the binary nature of T Tauri S was discovered almost double the
fraction of the orbit that has been observed so far.

%
% orbitNS.tex
% created by lmfinderrq on Wed Feb 20 17:59:58 2008
%
\begin{table}
\caption{Parameters of the best orbital solution for N -- S.}
\label{OrbNSTab}
%% stretch a bit so that sub- and superscripts don't overlap
\renewcommand{\arraystretch}{1.3}
\begin{center}
\begin{tabular}{lcr@{}l}
\noalign{\vskip1pt\hrule\vskip1pt}
Orbital Element				&  & \multicolumn{2}{c}{Value} \\
\noalign{\vskip1pt\hrule\vskip1pt}
Date of periastron $T_0$			&  & $2452981$ & $\,^{+8700}_{-3900}$\\
						&  & (2003 Dec  7)\span\\
Period $P$ (years)				&  & $  23240$ & $\,^{+11000}_{-12000}$\\
Semi-major axis $a$ (mas)			&  & $   9837$ & $\,^{+18000}_{ -5200}$\\
Semi-major axis $a$ (AU)			&  & $   1443$ & $\,^{+2640}_{ -762}$\\
Eccentricity $e$				&  & $   0.93$ & $\,^{+0.05}_{-0.05}$\\
Argument of periastron $\omega$ ($^\circ$)	&  & $ 198$ & $\,^{+ 7}_{ -5}$\\
P.A. of ascending node $\Omega$ ($^\circ$)	&  & $ 351$ & $\,^{+14}_{-11}$\strut\\
Inclination $i$ ($^\circ$)			&  & $  43$ & $\,^{+ 4}_{ -8}$\\
System mass $M_S$ ($M_\odot$)			&  & $   5.57$ & $\,^{+0.11}_{-0.29}$\\
%%$\chi^2$	       				&  & $  213.4$	\\
reduced $\chi^2$				&  & $    3.3$\\
\noalign{\vskip1pt\hrule\vskip1pt}
\end{tabular}
\end{center}
\end{table}

\begin{figure*}[ht]
\centerline{%
  \includegraphics[angle=90,width=.5\hsize]{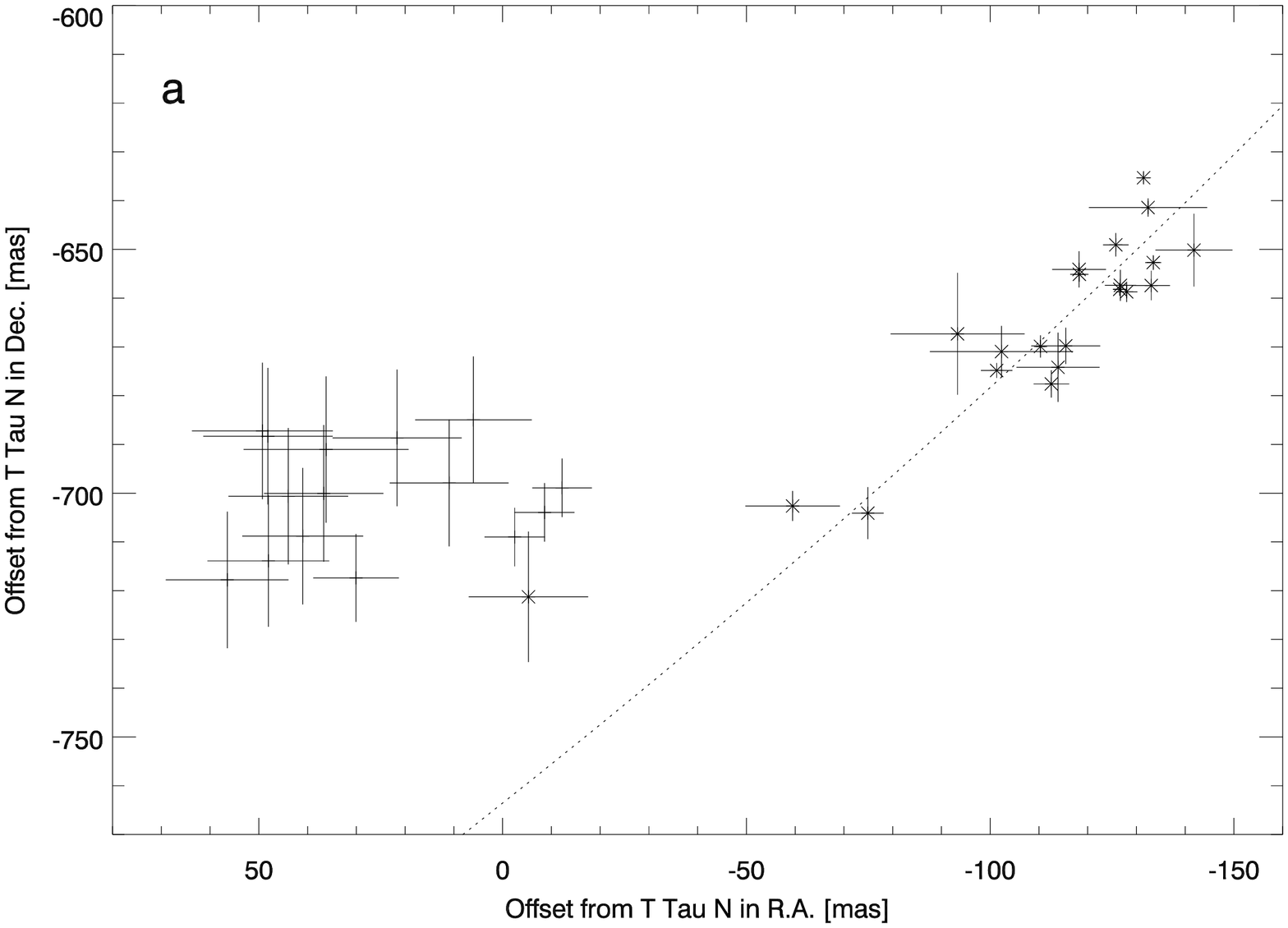}%
  \includegraphics[angle=90,width=.5\hsize]{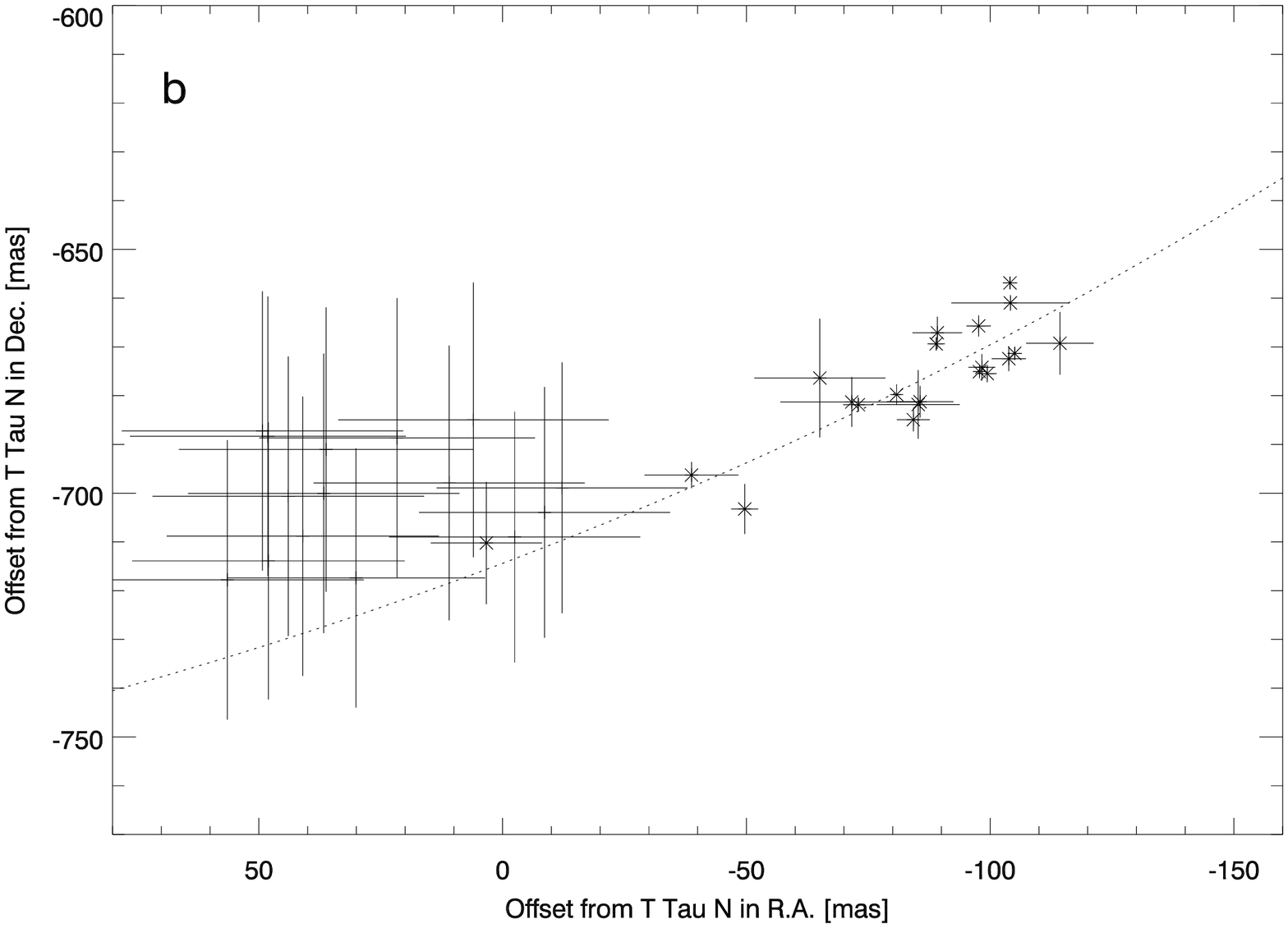}}
\centerline{%
  \includegraphics[angle=90,width=.5\hsize]{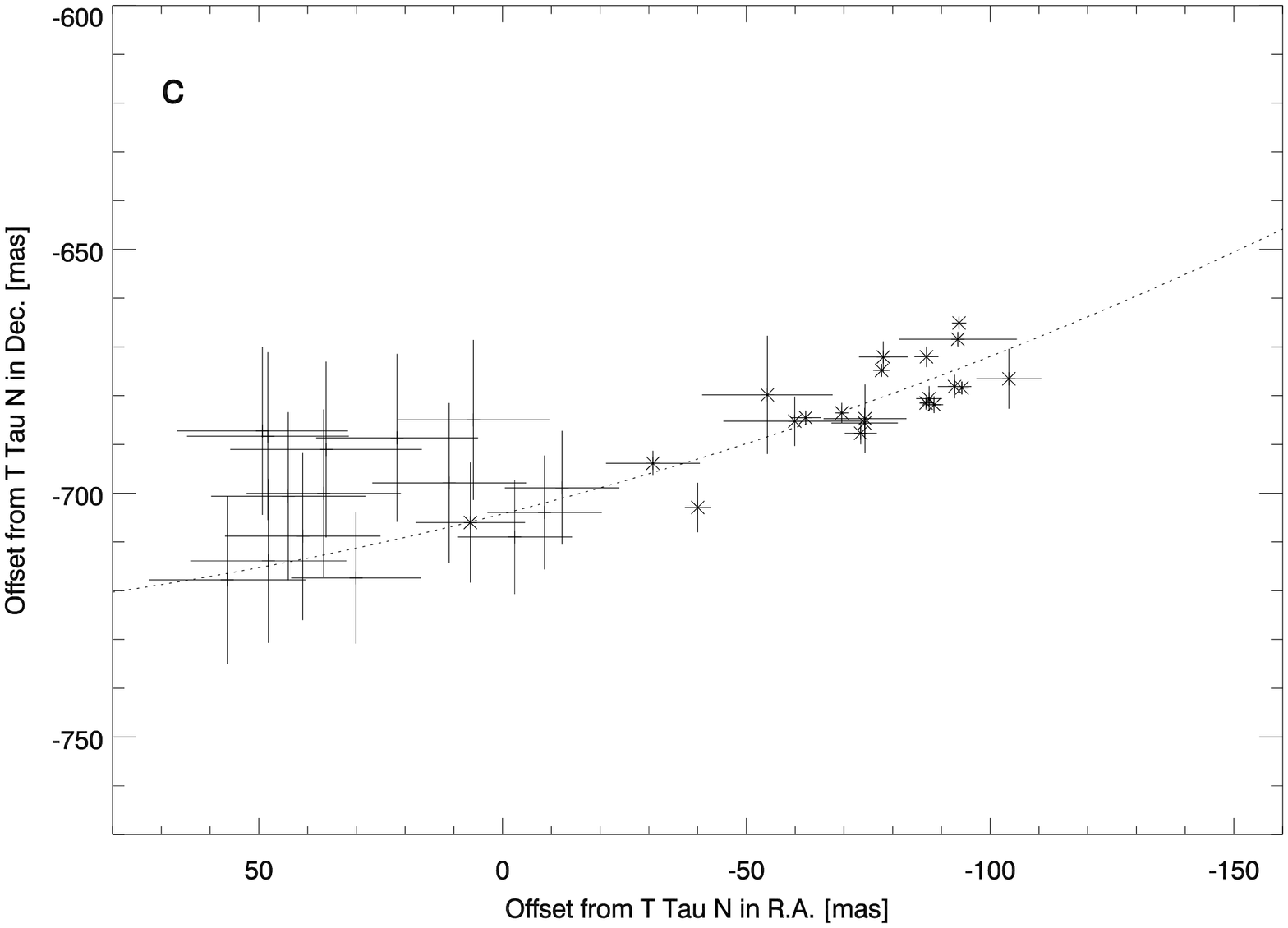}%
  \includegraphics[angle=90,width=.5\hsize]{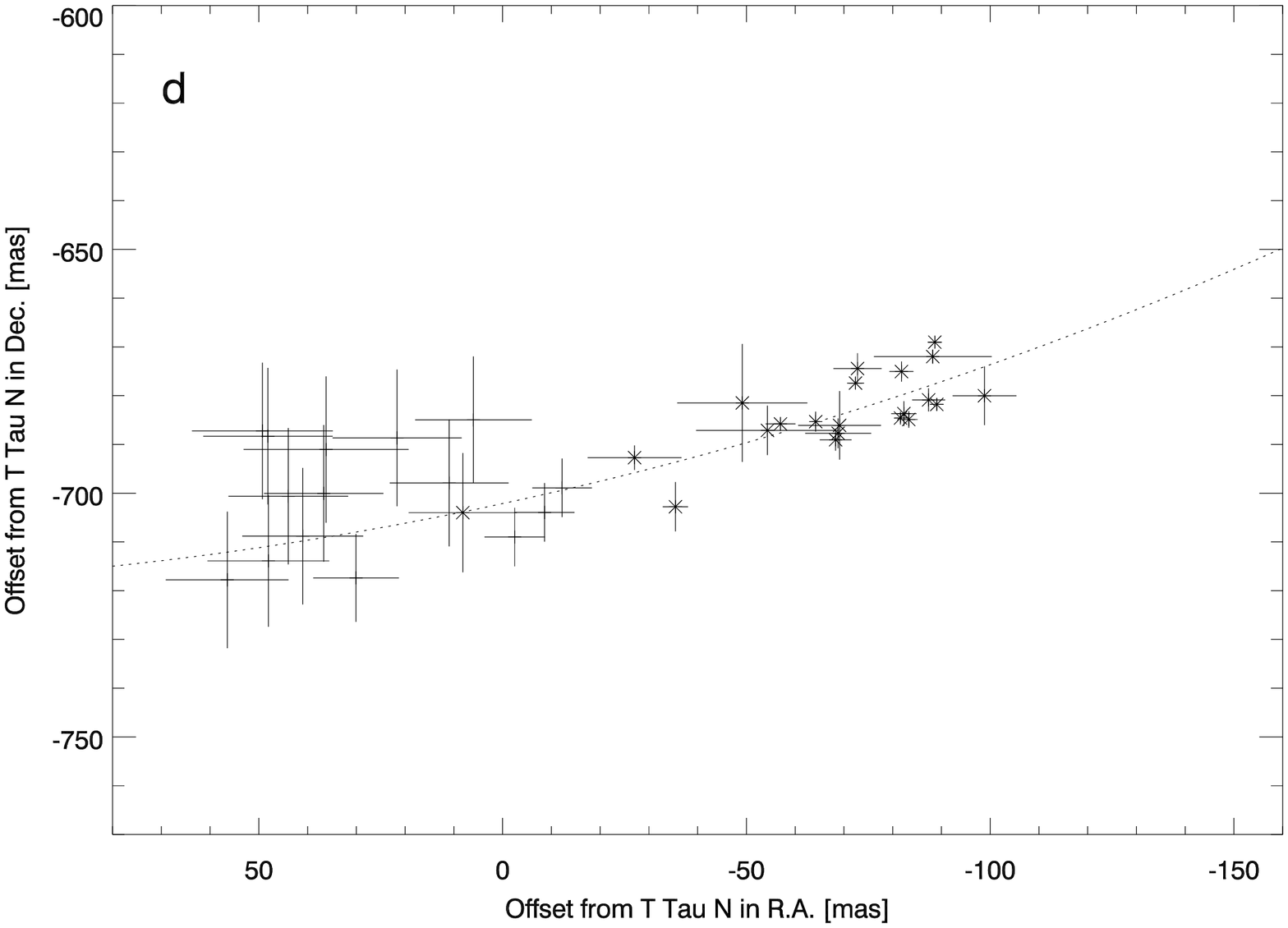}}
\caption{Results of fits for the orbit of the center of mass of
  T~Tau~S around T~Tau~N.  The four panels differ in the way
  observations were treated that did not resolve T Tau S into Sa and
  Sb.  These measurements are marked by crosses, measurements that did
  resolve T~Tau~S are marked by asterisks.  Note that the crosses show
  the center of light (which can be observed directly), while the
  asterisks show the center of mass (which depends on the observations
  and the model-parameter $q$).
  Panel a shows the result if unresolved observations are not used at
  all in the fitting procedure.  The unresolved measurements are
  plotted for comparison only.
  The orbit in panel b was obtained by adding 25\,mas uncertainty to
  the unresolved observations, while panel c shows the result if
  10\,mas are added.  For Panel d, the uncertainties given in the
  literature were used, without any additional errors to compensate
  for the offset between center of mass and center of light.  In all
  panels, the dotted line shows the path of the center of mass of T
  Tau S in our models.}
\label{NSorbits}
\end{figure*}

A fit only to observations that did resolve T Tau S results in an
orbit model that does not reproduce the positions measured before 1995
(Fig.~\ref{NSorbits}a).
Although we cannot derive the exact position of the center of mass, it
is unlikely that the observed center of light is as far off as this
model suggests.  Furthermore, the mass ratio this fit yields is 2.42,
i.e.\ according to the model, T~Tau~Sb would be 2.42 times more
massive than T~Tau~Sa.  With a system mass of $3\rm\,M_\odot$
(see section~\ref{SabFitSect}), the mass of T Tau Sb would be
$2.1\rm\,M_\odot$, inconsistent with its spectral type of about M1.

A compromise between using all observations and only observations that
resolved T Tau S is to use all observations, but add some
additional uncertainty to the errors of the unresolved observations,
in order to compensate for the offset between center of light and
center of mass.  Since the separation between T Tau Sa and Sb was
about 50\,mas in 1997, 25\,mas is a reasonable upper limit for the
separation between the two centers.  In principle, we
could use the predicted position angle of the Sa-Sb vector to
constrain the direction of the additional uncertainty.  However, the
predicted position angle depends strongly on the period of the orbit
of Sa and Sb around each other, which is rather uncertain.
Therefore, we decided to not apply any constraint on the
position angle of the vector from center of light to center of mass,
we simply add the same uncertainty in both $x$ and $y$.

Figure~\ref{NSorbits}b shows the result if 25\,mas are added in both
$x$- and $y$-direction.  The resulting orbit model is closer to the
unresolved observations than the model of Fig.~\ref{NSorbits}a, within
less than $2\sigma$ of the enlarged uncertainties.  The mass ratio is
0.77, so T Tau Sb should have a mass of about $1.3\rm\,M_\odot$,
which is still inconsistent with spectral type M1.

Figure~\ref{NSorbits}c shows the result of a fit where we added 10\,mas
additional uncertainty to the unresolved observations.  The mass ratio
of the best-fitting model is about 0.49, in which case the mass of
T~Tau~Sb is $1\rm\,M_\odot$.

\begin{figure}[ht]
\centerline{\includegraphics[angle=90,width=\hsize]{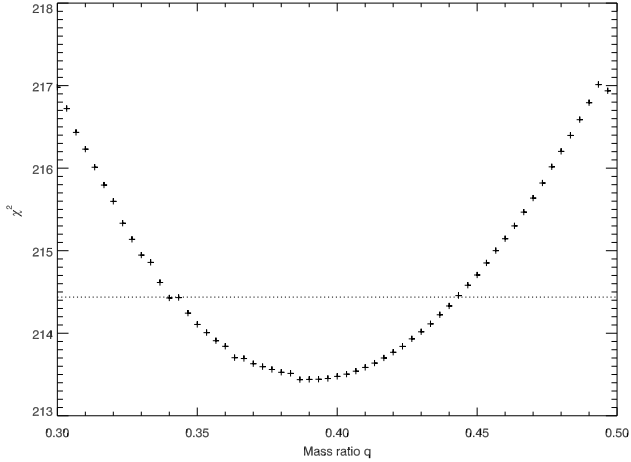}}
\caption{Minimum $\chi^2$ as function of the mass ratio $M_{\rm Sb} /
  M_{\rm Sa}$.  The dotted line marks $\chi^2_{\rm min}+1$, which
  defines the confidence interval for $q$.}
\label{chiminvsq}
\end{figure}

Finally, Fig.~\ref{NSorbits}d shows the result if we do not add any
additional uncertainties, i.e.\ take all the measurements and their
errors as they were published.  The mass ratio is $0.39\pm0.06$.  The
uncertainty of the mass ratio was estimated by finding the minimum and
maximum $q$ that yields a $\chi^2$ less than $\chi^2_{\rm min}+1$
(Fig.~\ref{chiminvsq}).  The $1\sigma$-uncertainty for $q$ is half the
width of this interval.  With this mass ratio, T~Tau~Sb would have a
mass of $0.83^{+0.10}_{-0.11}\rm\,M_\odot$, which would be in
reasonable agreement with its spectral type.
{\new Table~\ref{OrbNSTab} lists the orbital elements of the best fit,
  together with estimates for their confidence limits.  These
  confidence limits give the size of the region around the global
  minimum where $\chi^2 < \chi^2_{\rm min}+1$.  We did not search for
  local minima elsewhere in parameter space.  It is therefore possible
  that there are orbits with, e.g., significantly different
  inclination, which also fulfil the $\chi^2 < \chi^2_{\rm min}+1$
  criterion.  In particular, $\chi^2$ as function of period or
  semi-major axis is rather flat and appears to be dominated by random
  fluctuations, making it difficult to find a meaningful confidence
  limit.  The orbital elements given in Table~\ref{OrbNSTab} should be
  good enough to predict the positions up to a few decades in
  the past or future (depending on the required accuracy).}

We conclude that the mass ratio $M_{\rm Sb}/M_{\rm Sa}$ is probably in
the range 0.3 -- 0.5, but the time-span of observations that resolve
the Sa-Sb-pair is not long enough to exclude higher mass ratios.

\begin{figure}[t]
\centerline{\includegraphics[angle=90,width=\hsize]{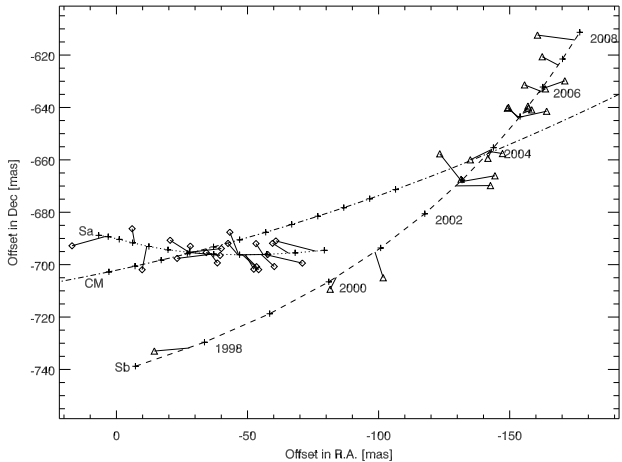}}
\caption{Motion of T Tau Sa and Sb in the reference frame of T~Tau~N.
The path of Sa predicted by our orbit models is shown by the dotted
line, the path of Sb by the dashed line, and the path of their center
of mass (CM) by the dash-dotted line.  The predicted positions on
January 1, 1997 to 2008 are marked by crosses.  The observed positions
of Sa are depicted by diamonds, those of Sb by triangles.  The solid
lines connect the observed and the predicted positions.}
\label{OrbNSabPic}
\end{figure}

Figure~\ref{OrbNSabPic} shows the motion of T~Tau Sa and Sb in the rest
frame of T~Tau N, including both the observed positions and those
predicted by our fits.

%%%%%%%%%%%%%%%%%%%%%%%%%%%%%%%%%%%%%%%%%%%%%%%%%%%%%%%%%%%%%%%%%%%%%%%%%%%%%

\section{Discussion}
\label{discuss}

\subsection{How realistic is our model?}
\label{modeldiscuss}

Our orbit model assumes that the center of mass of T Tauri S is on a
true Keplerian orbit, which is not disturbed by the binary nature of
the southern component.  Whether this assumption is valid is difficult
to judge with the data at hand, since it depends on the minimum
distance of Sa and Sb to N, which in turn depends on the shape and
orientation of both orbits.

Our best fit for the orbit of T Tau Sa and Sb around each other has a
maximum separation of about 46\,AU.  The projected separation between
T Tau N and S is currently about 100\,AU.  This would be close enough
to the separation of Sa and Sb to cause significant perturbations on
the orbit of Sa/Sb.  However, we know only the projected separation,
little is known about the 3-dimensional distance.  The best orbit
resulting from the fits described in Sect.~\ref{NSFitSect} has a
semi-major axis of about 1500\,AU, albeit with a {\newer high} eccentricity,
which leads to a periastron distance of 105\,AU.  In fact, the
semi-major axes of N-S orbits with a small $\chi^2$ span the range
from about 180\,AU to 1800\,AU, but all their periastron distances are
between 100 and 110\,AU.  This might indicate that the orbit of Sa and
Sb around each other is indeed perturbed by T Tau N, but we have to
wait for observations of a much larger fraction of the N-S orbit to
draw any conclusions.  With the data currently available, it is not
possible to detect deviations from a Kepler-orbit.

\subsection{Why is the mass so well constrained?}
\label{exactmasssec}

It is surprising that the system mass of T Tauri S derived from our
orbital fits has a much smaller fractional error than the semi-major
axis and the period.  Mathematically, this follows from the
correlation of $a$ and $P$ around the minimum $\chi^2$.  This
correlation is found as a result of the fitting procedure.  However,
there is also a physical reason for it.

Newton's law of gravitation states that the acceleration vector
$\ddot r$ of the companion is directed towards the primary star, and
that it is proportional to the system mass $M$ and the inverse of
their distance $r$ squared:
$$
	\ddot r = G M r^{-2},
$$
where $G$ is the constant of gravitation. If we could measure $r$ and
$\ddot r$ at any point of the orbit, we could compute $M$.  However,
we cannot observe the component of $r$ along the line of sight.  There
are only two points in the orbit where this component is zero: the
nodes of the orbit.  At
these points, we can observe $r$ and $\ddot r$ and
compute the mass. In fact, according to our best orbit model,
T~Tauri South passed through a node in 2002, in the middle of the
time interval covered by the observations.  In principle, we can use
the observations near this time to compute the mass directly from
$r$ and $\ddot r$.  In practice, uncertainties in the measurements
prevent us from obtaining a reliable result.
However, any orbit model approximating the measured positions will
also contain good approximations for $r$ and $\ddot r$, and therefore
lead to the same system mass.
This means that the system mass does not
vary much among orbit models approximating the observations.  A
consequence of this nearly constant mass is the correlation of $a$ and
$P$, since Kepler's third law states that $a^3$ is proportional to
$P^2$, if the mass is constant.

\subsection{Comparison to previous results}

\cite{duchene06} presented an orbital solution for the T Tau S binary
based on IR measurements published before 2006 and radio observations
at 2 and 3.7\,cm wavelength (they assume that the radio source is
identical to T~Tau~Sb).  They adopted the distance to T~Tau of
$141.5\pm2.8\rm\,pc$ estimated by \cite{loinard05}, while we use the
more recent value of $147.6\pm0.6\rm\,pc$ \citep{loinard07b}.
\cite{duchene06} find a period of $21.66\pm0.93\rm\,yrs$, a semi-major
axis of $82.1\pm1.8\rm\,mas$, and a total mass of T~Tau~S of
$3.79\pm0.41\,M_{\sun}$ (scaled to the more recent distance
determination).  This is inconsistent with our results, and they
estimate uncertainties that are almost two orders of magnitude smaller
than ours.  We believe that this is caused by the method they used to
derive the uncertainties.

\cite{duchene06} employed a bootstrap method to estimate the errors in
the fitted orbital elements.  They created artificial datasets where
the data points were randomly drawn from Gaussian distributions
centered on the actual measurements.  The $\chi^2$-minimization
routine was run on each of the artificial datasets, which results in a
distribution for each of the parameters.  The widths of these
distributions were used as an estimate for the errors of the fit
parameters.

This bootstrap method is not a good way to estimate errors in the case
of T Tauri S.  We have only observational data for about one quarter
of the orbit.  Therefore, a {\newer wide} range of, e.g.,
periods is compatible with the data (see Fig.~\ref{chiminfig}).
In the part of the orbit that has been observed so far, all these
orbits are close to the observed data points, and hence also close to
each other.  The orbits show very different separations in the parts
that have not (yet) been observed.  The uncertainty in the orbital
parameters is caused by incomplete coverage of the orbit, {\em not\/}
by the measurement errors of the observations.  Bootstrapping methods
help in determining how errors in the input data propagate to errors
in the results.  In the case of T Tauri, the measurement errors
contribute only a small part of the uncertainty of the orbital
elements. Therefore, bootstrapping methods underestimate the errors.

The best way to estimate confidence limits on the fitted parameters in
this case is to analyze $\chi^2$ as a function of the parameter values
(\citealt{press92}, see also Sect.~\ref{SabFitSect}).
The best estimate for the fitted parameters is, of course, the point
where $\chi^2$ is a minimum.
As confidence limit for each parameter, we use the interval within
which $\chi^2 - \chi^2_{\rm min} < 1$.  These limits contain 68\,\%
of the probability distribution.
A note of caution is indicated here: The orbital elements are most
certainly not normally distributed.  Therefore, a single $\sigma$ is
not sufficient to describe the distribution.
First, the confidence limits are not symmetric around the minimum
(cf.\ Table~\ref{OrbSabTab}).
Second, the interval containing 95\,\% of the probability distribution
is not twice as large as the interval containing 68\,\%.  The
well-known rules for normal distributions do not apply here.  The
95\,\% confidence interval is defined by $\chi^2-\chi^2_{\rm min}<4$,
but since $\chi^2$ is not a quadratic function of the orbital
parameters, the interval is not twice as large as the interval defined
by $\chi^2-\chi^2_{\rm min}<1$.
We determined only the 68\,\% confidence intervals and list them in
Table~\ref{OrbSabTab}, they should suffice to indicate the
uncertainties in the parameter estimates.

%%%%%%%%%%%%%%%%%%%%%%%%%%%%%%%%%%%%%%%%%%%%%%%%%%%%%%%%%%%%%%%%%%%%%%%%%%%%%

\section{Summary\label{sum}}

We have collected all available near-infrared astrometric data on the
T~Tauri system from the literature.  We also present a new analysis of
the data published in \cite{koehler00a} and 5 new data points obtained
with NACO at the VLT.  Binary orbit models were fitted to the relative
positions of T~Tau Sa and Sb in order to estimate the orbital
elements.  We find that most elements are not very well constrained,
e.g.\ the period is $93^{+115}_{-55}\rm\,years$.  The total mass of
the T Tau S binary, however, can be estimated more precisely to be
$3.0^{+0.15}_{-0.24}\rm\,M_\odot$.
{\new It is unlikely that T Tau Sb was recently ejected from the
  system and is now in a highly eccentric orbit or even escaping from
  the system.  However, some orbit solutions with $e > 1$ are within
  $1\,\sigma$ of the best fit, so an unbound orbit cannot be ruled out
  completely.}

We used the positions of T Tau Sa and Sb relative to T Tau N to fit a
combined model of the N-S and the Sa-Sb binary. The separation of Sa
and Sb is taken from the observational data, only the mass ratio
$M_{\rm Sb}/M_{\rm Sa}$ is a free parameter for the fit.  Due to the
small fraction of the N-S orbit observed so far, no reliable
constraints for the orbital elements of the N-S binary could be
obtained from the data, but the mass ratio of the southern binary can
be estimated to be $0.4\pm0.1$.  This corresponds to individual
mass estimates for T~Tau Sa and Sb of
$2.1\pm0.2$ and $0.8\pm0.1\,M_\odot$.
The spectral type of T Tau Sb was used to decide which orbital
solution is acceptable, it is therefore not surprising that the mass
estimate is in agreement with the spectral type.

Our results indicate that T~Tau Sa is at least as massive as T Tau
N, although it is much fainter at optical wavelengths.  This can be
explained by extinction due to circumstellar (or circumbinary)
material, possibly an edge-on disk.  Observations at longer
wavelengths with even higher spatial resolution are necessary to study
this material (Ratzka et al., in prep.).

%%%%%%%%%%%%%%%%%%%%%%%%%%%%%%%%%%%%%%%%%%%%%%%%%%%%%%%%%%%%%%%%%%%%%%%%%%%%%

\acknowledgements

The discussions with Sabine Reffert and Nick Elias provided valuable
input for this work.  We also wish to thank the anonymous referee for
his or her report.

\bibliographystyle{bibtex/aa}
\bibliography{TTauOrbit}

\end{document}